\documentclass[pre,twocolumn,showkeys,superscriptaddress,nofootinbib]{revtex4-1}
\usepackage{amsmath}
\usepackage{amsfonts}
\usepackage[colorlinks,urlcolor=blue,linkcolor=blue]{hyperref} 
\usepackage{graphicx}
\usepackage[caption=false]{subfig}

\begin{document}
\title{Spontaneous Circulation of Active Microtubules Confined by Optical Traps} 

\author{Stephen E Martin}
\affiliation{Department of Physics, University of California, Santa Cruz, CA 95064, USA}
\author{Matthew E Brunner}
\affiliation{Voltaiq Inc. 2150 Shattuck Ave, \#704 Berkeley, CA 94704}
\author{Joshua M Deutsch}
\email{josh@ucsc.edu}
\affiliation{Department of Physics, University of California, Santa Cruz, CA 95064, USA}
\date{\today}

\begin{abstract}
\noindent
We propose an experiment to demonstrate spontaneous ordering and symmetry breaking of kinesin-driven microtubules confined to an optical trap. Calculations involving the feasibility of such an experiment are first performed which analyze the power needed to confine microtubules and address heating concerns. We then present the results of first-principles simulations of active microtubules confined in such a trap and analyze the types of motion observed by the microtubules as well as the velocity of the surrounding fluid, both near the trap and in the far-field. We find three distinct phases characterized by breaking of distinct symmetries and also analyze the power spectrum of the angular momenta of polymers to further quantify the differences between these phases. Under the correct conditions, microtubules were found to spontaneously align with one another and circle the trap in one direction.
\end{abstract}

\maketitle


\section{Introduction}

The study of active fluids -- fluids which have a source of internal propulsion, as is common in biology -- has been of interest for several decades. In more recent years, several experiments have demonstrated remarkable emergent phenomena using mixtures of microtubules and kinesin fueled by ATP. Kinesin is a motor protein that binds to a microtubule in such a way that, when powered by ATP, tends to "walk" along the microtubule in a particular direction governed by the microtubule polarity. As the kinesin moves, it feels a drag force by the surrounding fluid. This has the effect of pushing the fluid in the direction that the kinesin is moving, as well as (by Newton's third law) exerting a force on the microtubule in the opposite direction. As such, flows can be observed in the fluid with no external impetus, from vortex lattices\cite{sumino2012large} to 2-dimensional active nematics\cite{sanchez2012spontaneous,decamp2015orientational} to macroscopic coherent flow\cite{wu2017transition}.

In what follows, we use first-principles hydrodynamic simulations to show the types of behavior (e.g. spontaneous circulation) of confined microtubules. We propose an experiment in which microtubules are held in place by an optical trap. To aid in this, we show the feasibility of such an experiment by first calculating the laser power required to contain a bending microtubule, and then we derive the change in temperature one might expect for such an experiment.

We then discuss the methods of simulation. While the specifics of the simulation are somewhat complicated, it uses no phenomenology. In fact, it has (in slightly adapted forms) been used to successfully reproduce and offer important insights into the phenomena of cytoplasmic streaming in Drosophila oocytes\cite{Monteith2016} and metachronal wave formation in microtubule bundles\cite{martin2018emergence}.

The results of the simulations themselves are then presented, in which we identify qualitatively different kinds of observed motion. We examine the dependence of these outcomes on parameters, and provide some interpretation. In addition to examining the motion of the polymers themselves, we also calculate and present analysis of the fluid motion in the near- and far-field. We then perform a power spectrum analysis of the angular momenta of polymers to give some concrete metrics for determining phase.

\section{Preliminary Estimations}

Before going in-depth into predictions as to what will happen if microtubules are contained in an optical trap, we should first address whether it is feasible to do so given typical issues common in optical trapping.

\subsection{Laser frequency}

For biological applications, lasers with wavelength $> 1000$ nm are typical, as this reduces damage to cells \cite{ashkin1987cell, stevenson2006optically}. However, even from the beginning of its usage, optical traps with wavelengths in the visible regime have been successfully used for sub-cellular structures \cite{ashkin1987virus}. In what follows, we consider a range of wavelengths from 500 nm to slightly over 1000 nm.

\subsection{Forces}

Microtubules confined in an optical trap would bend, and it is useful to get a sense of whether an optical trap would exert a sufficient force to keep the microtubules contained. In what follows, we use typical values to argue that an optical trap with reasonable intensity would be enough to overcome the microtubule rigidity.

The most common and straightforward approach to optical trapping uses a tightly-focused Gaussian beam in the TEM$_{00}$ mode. For such a beam, the intensity of the radiation as a function of radial position can be written as
\begin{equation}
\label{eq:gaussian_I}
I(r) = \frac{2P}{\pi w^2} \exp\left(-\frac{2r^2}{w^2}\right)
\end{equation}
where $P$ is the power of the laser and $w$ is the radius of the beam cross section. Because the diameter of a microtubule is 24 nm and the smallest the beam diameter can be is on the order of the wavelength $\lambda \sim 500$ nm, we use the electric dipole approximation
\begin{equation}
\label{eq:gaussian_F}
\mathbf{F} = \frac12\alpha \nabla E^2
\end{equation}
where $\alpha$ is the induced dipole of the trapped particle. For a sphere of radius $a$,
\begin{equation}
\label{eq:alpha}
\alpha = 4\pi n_0^2 \epsilon_0 a^3\frac{m^2-1}{m^2+2},
\end{equation}
where $m\equiv n_1/n_0$; $n_0$ and $n_1$ being the refractive indices of the surroundings and the sphere respectively\cite{harada1996radiation}. Because, for a monochromatic wave, $I = \frac{c\epsilon_0 n_0}{2}E^2$, we rewrite \ref{eq:gaussian_F} as
\begin{equation}
\label{eq:FwithI}
\mathbf{F} = \frac{4\pi n_0 a^3}{c}\frac{m^2-1}{m^2+2} \nabla I
\end{equation} 
In terms of the Gaussian beam described by equation \ref{eq:gaussian_I},
\begin{equation}
\label{eq:Fofr}
F(r) = -\frac{32 n_0 a^3 P r}{c w^4}\left(\frac{m^2-1}{m^2+2}\right)\exp\left(-\frac{2r^2}{w^2}\right) 
\end{equation}
For a bending microtubule, the elastic restoring force per unit length will be
\begin{equation}
\label{eq:fel}
f_{el} = C\left|\frac{d^4\mathbf{r}}{ds^4}\right|
\end{equation}
If the microtubule is circling the optical trap, then this becomes
\begin{equation}
\label{eq:elasticf}
f_{el}(r) = \frac{C}{r^3}
\end{equation}
Where $r$ is the radius of circulation. If we now model the microtubule as a chain of beads (each bead having radius $a$), then the elastic force on a single bead would be
\begin{equation}
\label{eq:elasticF}
F_{el}(r) = \frac{2aC}{r^3}
\end{equation}
Equating Eqs. \ref{eq:elasticF} and \ref{eq:Fofr} and solving for $P$ gives the power required to contain a microtubule circling at radius $r$,
\begin{equation}
\label{eq:Pofr}
P(r) = \frac{cC w^4}{16n_0 a^2r^4}\left(\frac{m^2+2}{m^2-1}\right)\exp\left(\frac{2r^2}{w^2}\right)
\end{equation}
This is minimized at $r=w$, meaning the minimum power required to contain a microtubule is 
\begin{equation}
\label{eq:Pmin}
P_{min} = \frac{cCe^2}{16n_0 a^2}\left(\frac{m^2+2}{m^2-1}\right)
\end{equation}
The microtubule stiffness $C$ has be measured to be approximately $1\times10^{-23}$ Nm$^2$\cite{gittes1993flexural,felgner1996flexural}. The refractive index of water is $n_0 = 1.33$, and the refractive index of tubulin has been measured to be $n_1\approx 2.5$ \cite{bon2014fast,mershin2004tubulin}. Modeling the microtubule as a string of beads of radius $a$ means $a\approx 12$ nm. Using these quantities, we find that $P_{min} \approx 16$ W. This is quite large compared to standard lasers used for optical traps, but it should be emphasized that when a group of microtubules are circling an optical trap, each individual microtubule often has a radius of curvature larger than the radius of the trap if the length of the microtubule is on the order of, or less than, the trap radius (see images in Section \ref{sec:results}). The minimum power input $P_{min}\propto b^{-3}$ where $b$ is the polymer radius of curvature, so even a factor of 2 increase (i.e. let $b=2r$) reduces $P_{min}$ to only 2 W.

\subsection{Heating}

The relatively large laser power required to perform this proposed experiment raises some concerns about heating in the system, and it is worthwhile to address the degree of heating one might expect so that an experiment may be properly designed. It has been shown\cite{peterman2003laser} that for a Gaussian beam, the change in temperature at the center of the optical trap of uniform absorbance is approximately
\begin{equation}
\label{deltaT}
\Delta T(r=0) \approx \frac{\alpha P}{2\pi C}\left[\ln\left(\frac{2\pi R}{\lambda}\right)-1\right],
\end{equation}
where $\alpha$ is the absorption coefficient, $P$ is the laser power, $C$ is the thermal conductivity, $\lambda$ is the laser wavelength, and $R$ is a characteristic distance ($R\gg \lambda$) to a boundary at which temperature is held constant, often taken to be the distance to the glass slide in experiments. If the experiment is performed on water, it should be noted that $\alpha$ is highly dependent on $\lambda$, but this dependence has been thoroughly studied \cite{kedenburg2012linear}.

\begin{figure}
\begin{center}
\includegraphics[width = \columnwidth]{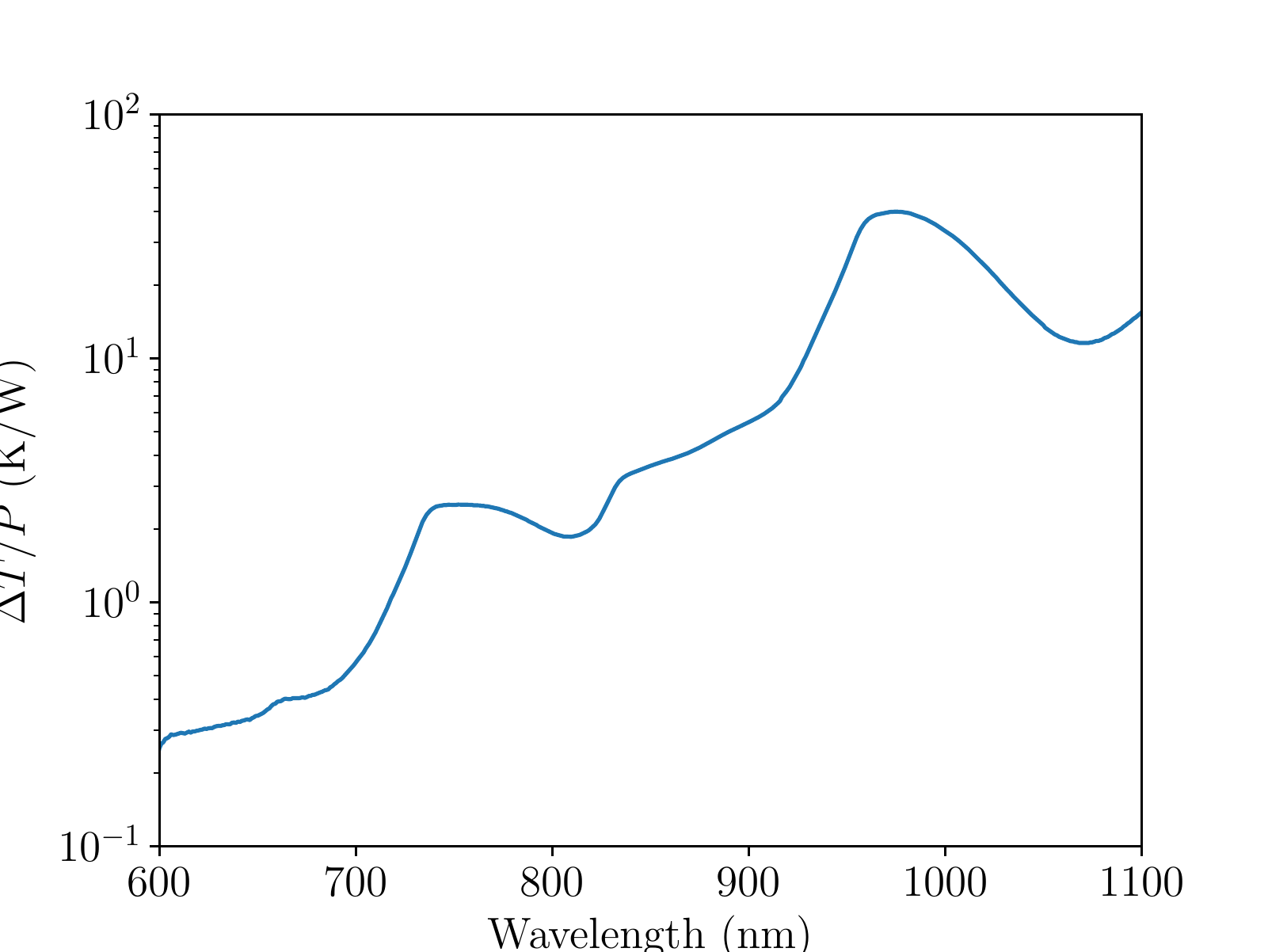}
\end{center}
\caption{Predicted increase in temperature per watt at the center of the optical trap as a function of laser wavelength. 
The trap is assumed to only contain water a distance of 10 $\mu$m from a thermally conducting plate. The lower bound of the plot terminates at 600 nm because at this point the absorption coefficient of water becomes similar to that of glass, and the constant 
temperature boundary condition approximation becomes questionable.
\label{fig:TvsWL}
}
\end{figure}

Fig. \ref{fig:TvsWL} shows $\Delta T(r=0)/P$ as a function of $\lambda$ for water, letting $R = 10\ \mu$m. From this, we can see that, for a 20 W laser, wavelengths longer than $\sim$700 nm quickly become unfeasible, as this would lead to temperature increases of 20 K or more. However, wavelengths shorter than 700 nm would likely only heat by 5-10 K.

Adding microtubules to the optical trap would increase the temperature further, as the absorbance of proteins tends to be several orders of magnitude larger than that of water. However, an important implication of the calculation in the previous section is that the trap radius has no effect on the power required to confine the microtubules. This means that, if needed, the radius could be increased (and microtubules made longer) in order to reduce the relative area of the microtubules. For example, a 1 $\mu$m long, 24 nm diameter microtubule takes nearly 1\% of the area of a trap of radius 1 $\mu$m. If the length of the microtubule and the radius of the trap are both increased to 10 $\mu$m, the area fraction per microtubule reduces by a factor of 10 to under 0.1\%.

Furthermore, the assumption that the optical trap is far from the plates will not necessarily be true, and therefore the calculations above should be seen as a upper bound on heating. Two plates separated by a very thin gap (on the order of or even less than the radius of the trap) may be used, as the hydrodynamics of this are accounted for in the simulations that follow. This would further increase the ability of the slides to dissipate heat, especially if the slide material is chosen to have high thermal conductivity and low absorbance at the desired wavelength. Sapphire substrates have been used for this purpose as it has $\sim20$ times the thermal conductivity than borosilicate glass \cite{polinkovsky2014ultrafast}, but quartz ($\sim3$ times the thermal conductivity of water) would likely also be a viable choice.

\section{Method of simulation}

This simulation was adapted from previous work used to successfully model cytoplasmic streaming and metachronal wave formation \cite{Monteith2016,martin2018emergence}, and more detailed explanations of what follows can be found in these papers. This is a first-principles simulation, and at a high level utilizes a straightforward approach. Microtubules are expressed as polymers, each made of a chain of $N$ monomers bound together by a spring force which separates them by a distance $\ell = 1$. A group of $M$ polymers is initialized, and various forces act on them. The $i$th monomer position and velocity are updated using a fourth-order Runge Kutte integration of the equation
\begin{equation}
\label{eq:RKint}
\frac{d\mathbf{r}_i}{dt} = \mathbf{u}(\mathbf{r}_i) - k_{kin}(\mathbf{r}_{i-1} - \mathbf{r}_{i+1})
\end{equation}
where $dt = 0.003$ is the time step, $k_{kin}=0.2$ corresponds to to the strength of the kinesin drag force tangent to the polymer (see following section), and $\mathbf{u}(\mathbf{r}_i)$ is the fluid velocity due to all other forces.

The polymers are initialized in the trap in a zig-zag pattern, alternating polarity, with random noise given to each monomer's initial position. This is more computationally efficient than true Monte Carlo initialization, and we observed no noticeable difference in simulation outcome.

\subsection{Kinesin drag force}

The drag force propelling the microtubules depends on the concentration of kinesin. Kinesin bind preferentially to microtubules, meaning that a relatively low concentration of kinesin in the solution will result in a high concentration of kinesin on the microtubules. Suppose that these kinesins are modeled as a linear train of spheres (radius $a$) separated by distance $d$ and moving at speed $v_0$. It has been shown \cite{Monteith2016} that, far from the kinesin, the fluid velocity is the same as that due to thin cylinder moving at speed $(a/d)v_0$. In terms of equation \ref{eq:RKint}, $k_{kin}\propto a/d$.

The above demonstrates that there are two main ways of tuning $k_{kin}$ experimentally: one can change the kinesin concentration (effectively changing $d$), or one can add cargo to the kinesin (changing $a$). Explicitly adding cargo for kinesin to transport is not strictly necessary -- many studies have shown active matter phenomena using no added cargo (although it is possible the kinesin are transporting segments of microtubule). This being said, the size of the dragged cargo is not particularly important in itself: a well-known result of slender-body theory is that the drag force $F$ on a cylinder moving parallel to its axis is

\begin{equation}
F \sim \frac{2\pi \mu \ell u}{\ln(\ell/a)},
\end{equation}

where $\mu$ is the dynamic viscosity, $u$ is the cylinder speed (relative to the far-field fluid), and $\ell$ is the cylinder length. This is only logarithmically dependent on the cylinder radius $a$. Much more important is the ratio $a/d$, as $u\propto a/d$ for the kinesin train.

\subsection{Description of other forces}

\begin{figure}
\centering
\subfloat[][]{
\includegraphics[width=0.3\columnwidth]{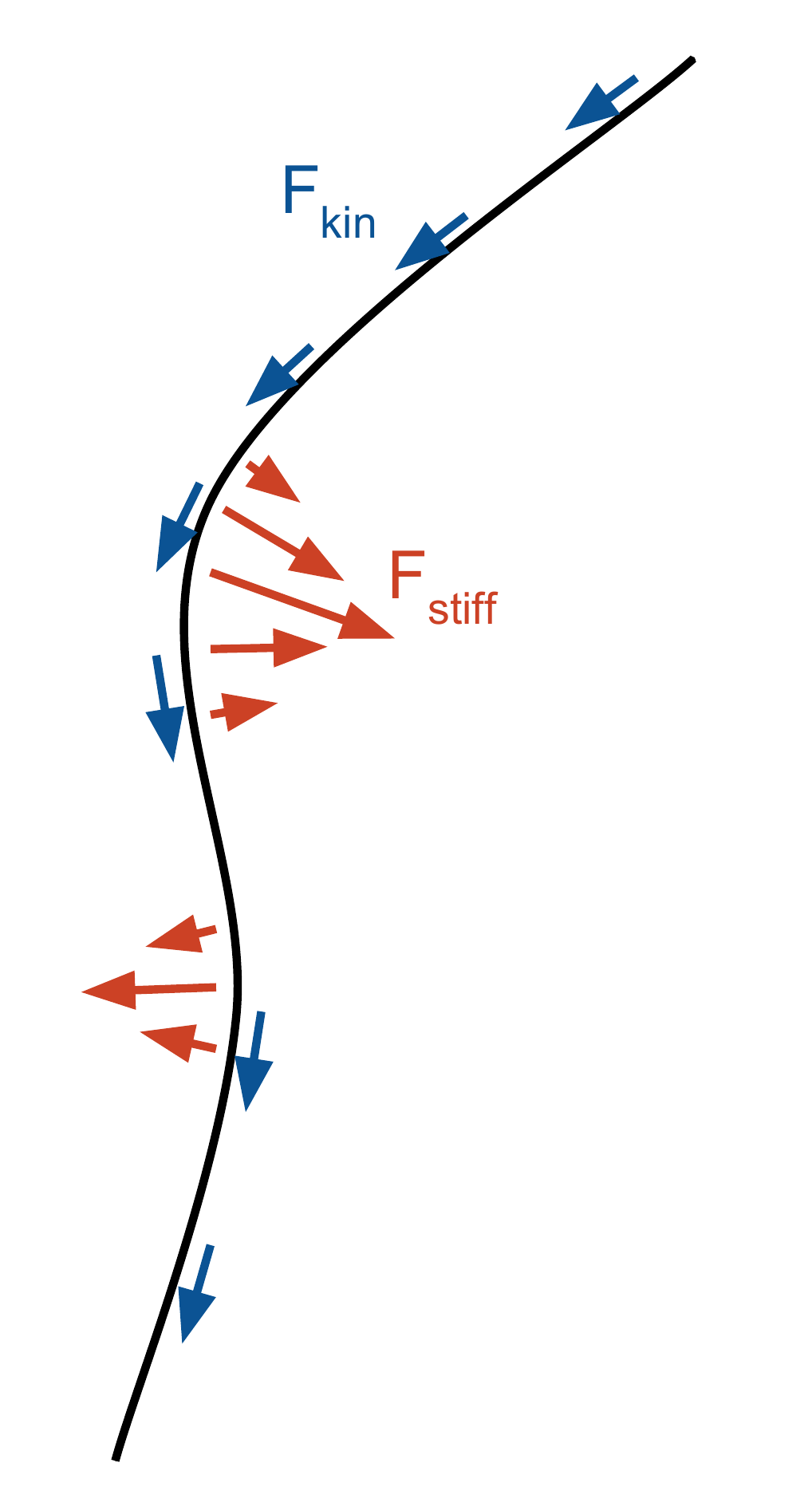}
}
\qquad
\subfloat[][]{
\includegraphics[width=0.4\columnwidth]{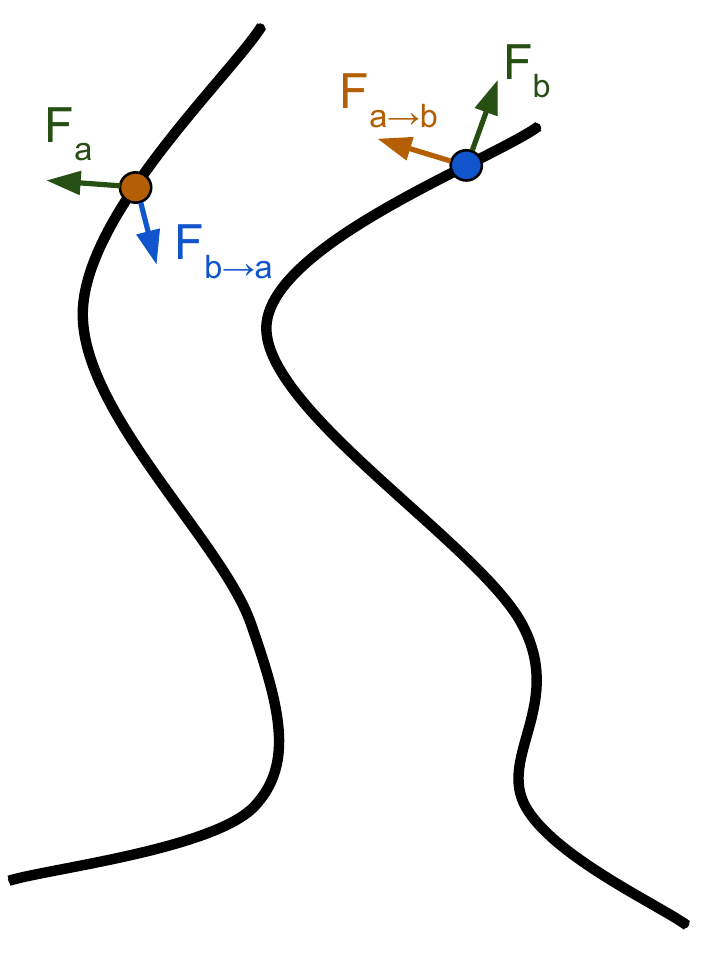}
}
\caption{Illustrations of some of the important forces applied to simulated polymers. (a) shows stiffness and kinesin drag forces that a polymer experiences regardless of the existence of other polymers (part of the $F$ terms in Eq. \ref{eq:uri}), and (b) gives an example of how polymers exert hydrodynamic forces on one another via the interaction tensor $\mathbb{G}$.\label{fig:forces}}
\end{figure}

The forces that go into computing $\mathbf{u}(\mathbf{r}_i)$ in equation \ref{eq:RKint} are:
\begin{equation}
\label{eq:uri}
\mathbf{u}(\mathbf{r}_i) = J_{ii}\mathbf{F}_i + \sum_{i\neq j}\mathbf{F}_j\cdot\mathbb{G}(\mathbf{r}_i - \mathbf{r}_j)
\end{equation}
Here, $J_{ii}\mathbf{F}_i = \mathbf{F}_i/4\pi\mu$ is the drag experienced by a small sphere, $\mathbb{G}$ is the hydrodynamic interaction tensor (described in the following section), and $\mathbf{F}$ is the sum of all forces on a monomer not related to kinesin drag or hydrodynamic interactions:
\begin{equation}
\label{eq:Fbreakdown}
\mathbf{F}_j = \mathbf{T}_j + \mathbf{C}_j + \mathbf{W}_j + \sum_k\mathbf{H}_{jk}
\end{equation}
where:
\begin{itemize}
\item $\mathbf{T}_j = k_{spr}\left[\left(|\mathbf{r}_{j-}|-\ell\right)\hat{\mathbf{r}}_{j-} + \left(|\mathbf{r}_{j+}|-\ell\right)\hat{\mathbf{r}}_{j+} \right]$\\
with $\mathbf{r}_{j\pm}\equiv \mathbf{r}_{j\pm 1} - \mathbf{r}_j$, is the spring force keeping monomer separation approximately constant. For our simulations, $k_{spr}=100$ and $\ell=1$.
\item $\mathbf{C}_j = k_{stiff}\left(2\textbf{r}_j - \textbf{r}_{j+2} - \textbf{r}_{j-2}\right)$\\
is the stiffness force which resists polymer bending. This is equivalent to $C$ in Eq. \ref{eq:fel}. $k_{stiff}$ is varied in our simulations, with $0.02 \leq k_{stiff} \leq 10$.
\item $\mathbf{W}_{j} = -k_{trap} \left|\frac{\mathbf{r}_j}{R}\right|^7\hat{\mathbf{r}}_j$\\
is the trapping force which pushes all monomers radially inward. The $r^7$ dependence of the trap was used for computational efficiency as well as to let polymers travel freely within the trap while providing a firm boundary at the trap radius $r=R$. A spring force of the form $\mathbf{W}_{j} = -k_{trap}\mathbf{r}_j$ was also attempted, and similar (albeit somewhat less stable) types of behavior were observed. This was not used for analysis, however, because the trap radius is less well defined and the time required for polymers to exhibit collective behavior is significantly longer. $k_{trap} = 1.0$ for all simulations, and the trap radius $R$ was varied between 2.5 and 10.0.
\item $\sum_k\mathbf{H}_{jk} = k_{rep}\left[1-\left(\frac{d_{rep}}{|\mathbf{r}_j - \mathbf{r}_k|}\right)^4\right]\left(\mathbf{r}_j - \mathbf{r}_k\right)$\\
\phantom{.}\hfill if $|\mathbf{r}_j-\mathbf{r}_k| < d_{rep}$\\
is a repulsive force between monomers that only acts if two monomers are very close to one another. For our simulations, we set $d_{rep}=0.5$ and $k_{rep} = 1.0$.
\end{itemize}

\subsection{Interaction tensor}

\begin{figure*}
\centering
\subfloat[][]{
\includegraphics[width = 0.29\textwidth]{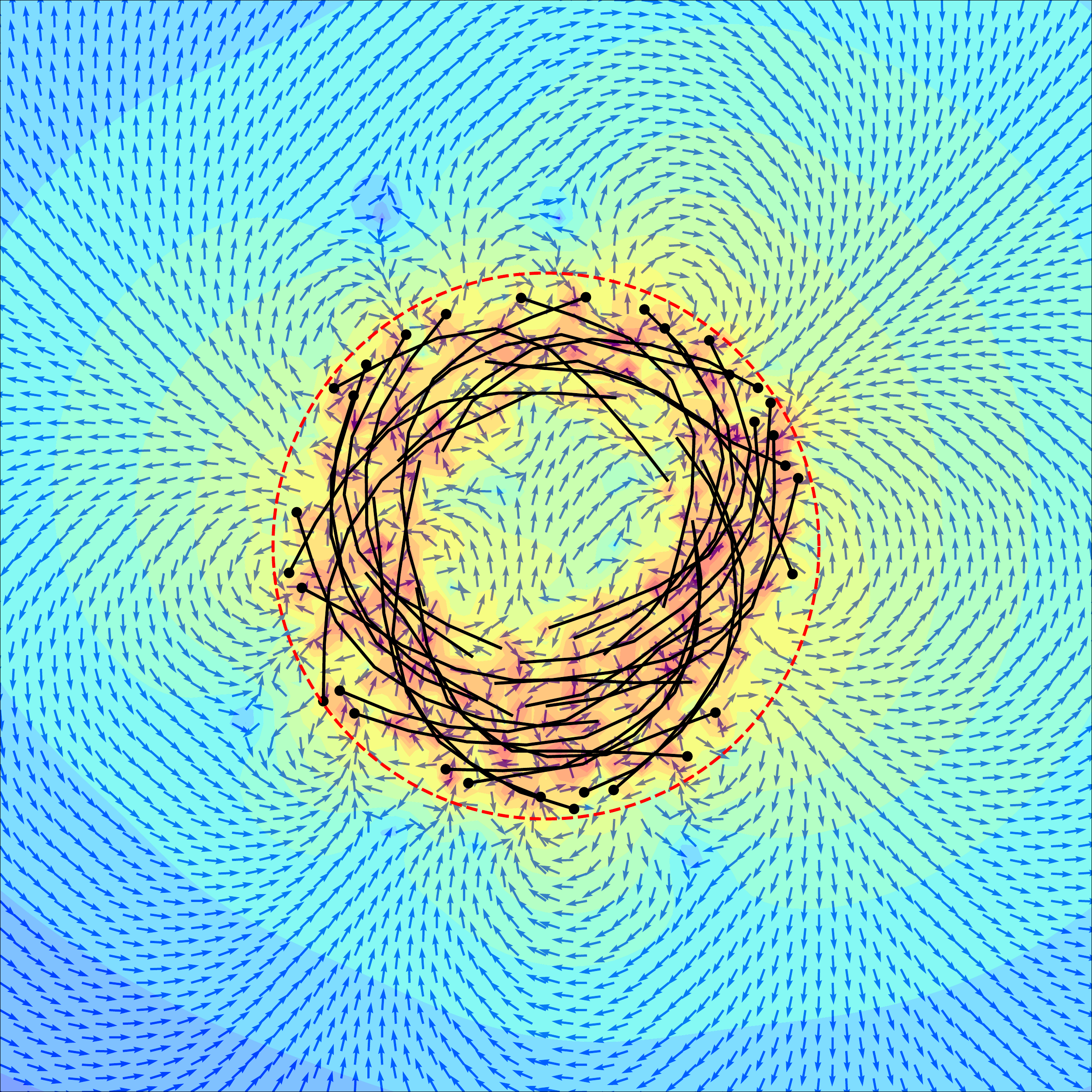}
}
\qquad
\subfloat[][]{
\includegraphics[width = 0.29\textwidth]{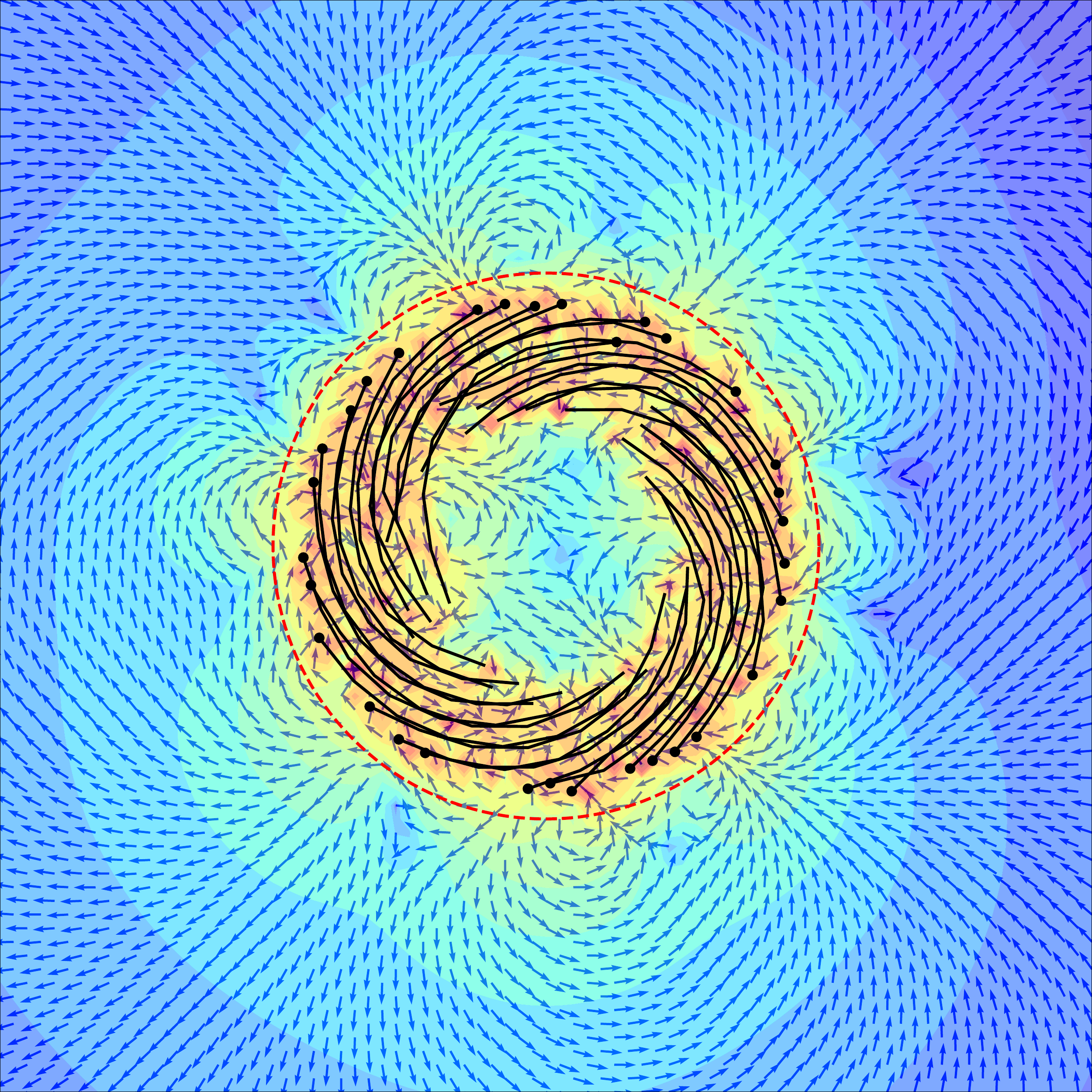}
}
\qquad
\subfloat[][]{
\includegraphics[width = 0.29\textwidth]{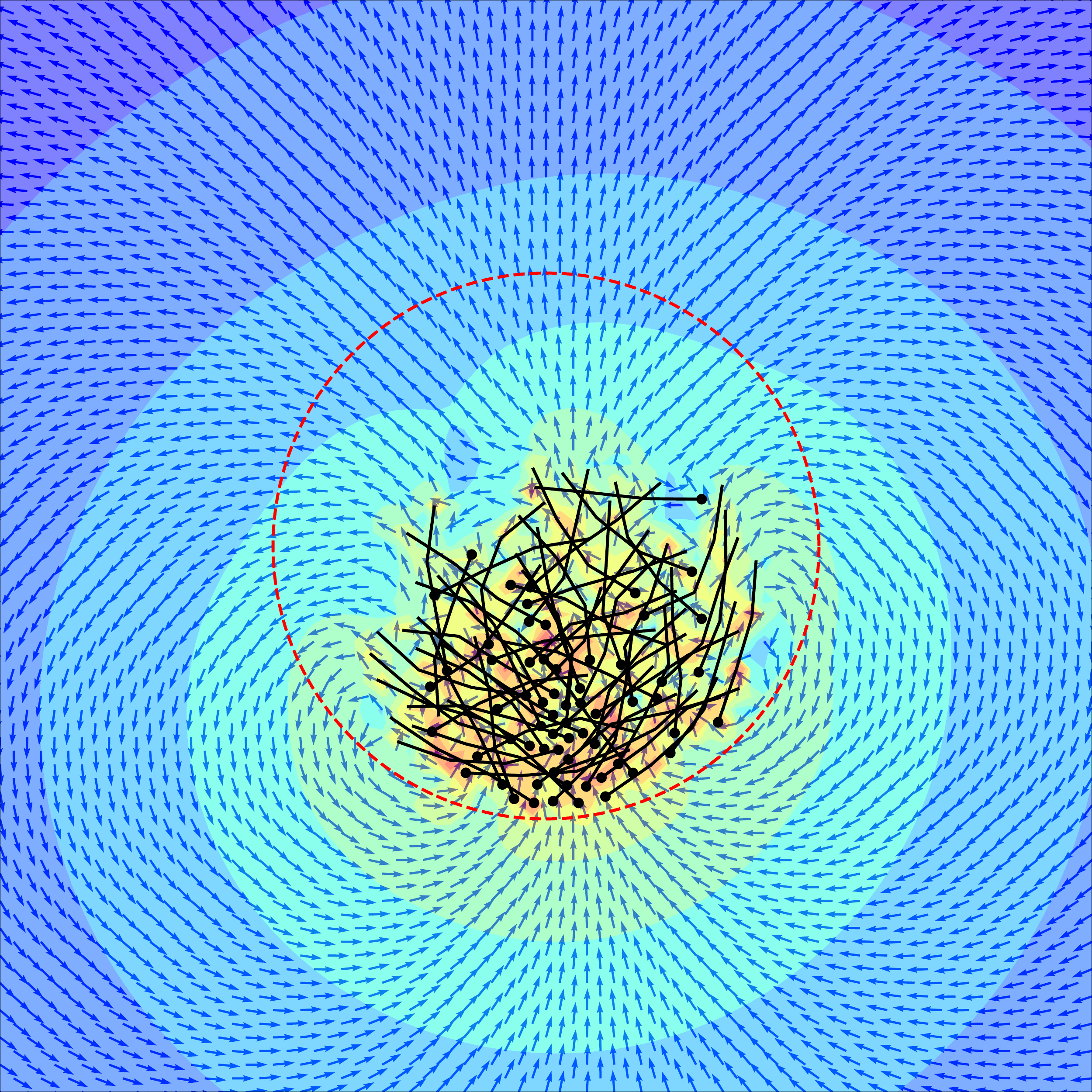}
}
\caption{Quiver plots showing the direction of fluid flow in the vicinity of the polymers (black curves, with black circles indicating polarity) caught in the trap (dashed red circle). Color map is overlaid to show relative magnitudes of fluid velocity. (a) Uncorrelated circulation, using 32 polymers of length 8 and $k_{stiff}=2.0$. (b) Correlated circulation, using 32 polymers of length 8 and $k_{stiff}=0.5$. (c) Stasis, with 64 polymers of length 4 and $k_{stiff}=0.5$. See appendix \ref{app:vid_links} for videos of each of these.\label{fig:quiver}}
\end{figure*}

The hydrodynamic interaction tensor $\mathbb{G}$ is the same as that which was used previously to simulate metachronal wave formation\cite{martin2018emergence}, so we will only describe its significance at a high level here. As this is Stokes flow (Re = 0), the sum of non-hydrodynamic forces exerted on each monomer is transferred perfectly to the surrounding fluid. If the monomer is sufficiently small relative to the interaction distances, such a force can be modeled as a point force (known as a stokeslet). The exact solution for the velocity field $\mathbf{v}(\mathbf{r})$ due to a free stokeslet as derived by Oseen has been known for nearly a century:
\begin{equation}
\label{eq:oseen}
\mathbf{v}(\mathbf{r}) = \frac{1}{8\pi\mu}\mathbf{F}\cdot \mathbb{J},
\end{equation} 
where
\begin{equation}
\label{eq:oseentensor}
\mathbb{J} \equiv \frac1r \left(\mathbb I + \frac{\mathbf{r}\otimes\mathbf{r}}{r^2}\right)
\end{equation}
is known as the Oseen tensor. 

The interaction tensor used in this study is a simplified version of the solution derived by Liron and Mochon for a stokeslet between two infinite flat parallel plates\cite{Liron1976,martin2018emergence}. We assume that microtubules in the proposed optical trapping experiment would be confined approximately halfway between two glass slides (separated by distance $H=1.0$). For this reason and for computational efficiency, we confine all monomers to this plane in all simulations.

\section{Results}
\label{sec:results}

In what follows, we describe the sorts of behaviors that emerge depending on input parameters. A link to videos of simulated behavior is provided in appendix \ref{app:vid_links}. We then analyze the velocity in the surrounding fluid to give a sense of the mixing ability of each type of behavior.

\subsection{Types of microtubule motion}

While the motion of simulated polymers was often complex, we classify behavior into three categories: uncorrelated circulation, correlated circulation, and stasis. These are broad classifications: the descriptions provided are qualitative, and many parameter choices exhibit intermediate behavior. Below is a description of these categories, and appendix \ref{app:categories} gives further information regarding how certain parameters affect the behavior exhibited by the system.

\subsubsection{Uncorrelated circulation}

If polymers do not interact sufficiently (i.e. if $k_{Oseen}$ is small) or polymer density is low, then each polymer tends to act independently of other polymers. As such, there is no symmetry breaking: approximately the same number of polymers circulate clockwise as counter-clockwise.  Correspondingly, the polarity of the microtubules is mixed.

\subsubsection{Correlated circulation}

If $k_{Oseen}$ is increased and the density of polymers is sufficient, then correlated circulation is often seen: that is, polymers interact strongly enough and the stiffness is low enough such that some polymers reverse direction, breaking symmetry to exhibit organized circulation in a single direction. The polarity of the microtubules are all in the same direction, consistent with the circulation of the system.

\subsubsection{Stasis}

If polymers interact strongly but the polymer length is short compared to the trap diameter, the stiffness is too low, or the polymer density is too high, then no circulation occurs. Polymers interact but are unable to change direction, resulting in a cluster of polymers that remains largely stationary, with occasional and irregular changes in direction. Here static rotational symmetry is broken, with the microtubules clustered in one region of the trap away from the middle.

\subsection{Fluid velocity field}

\begin{figure}
{
\centering
\includegraphics[width=\columnwidth]{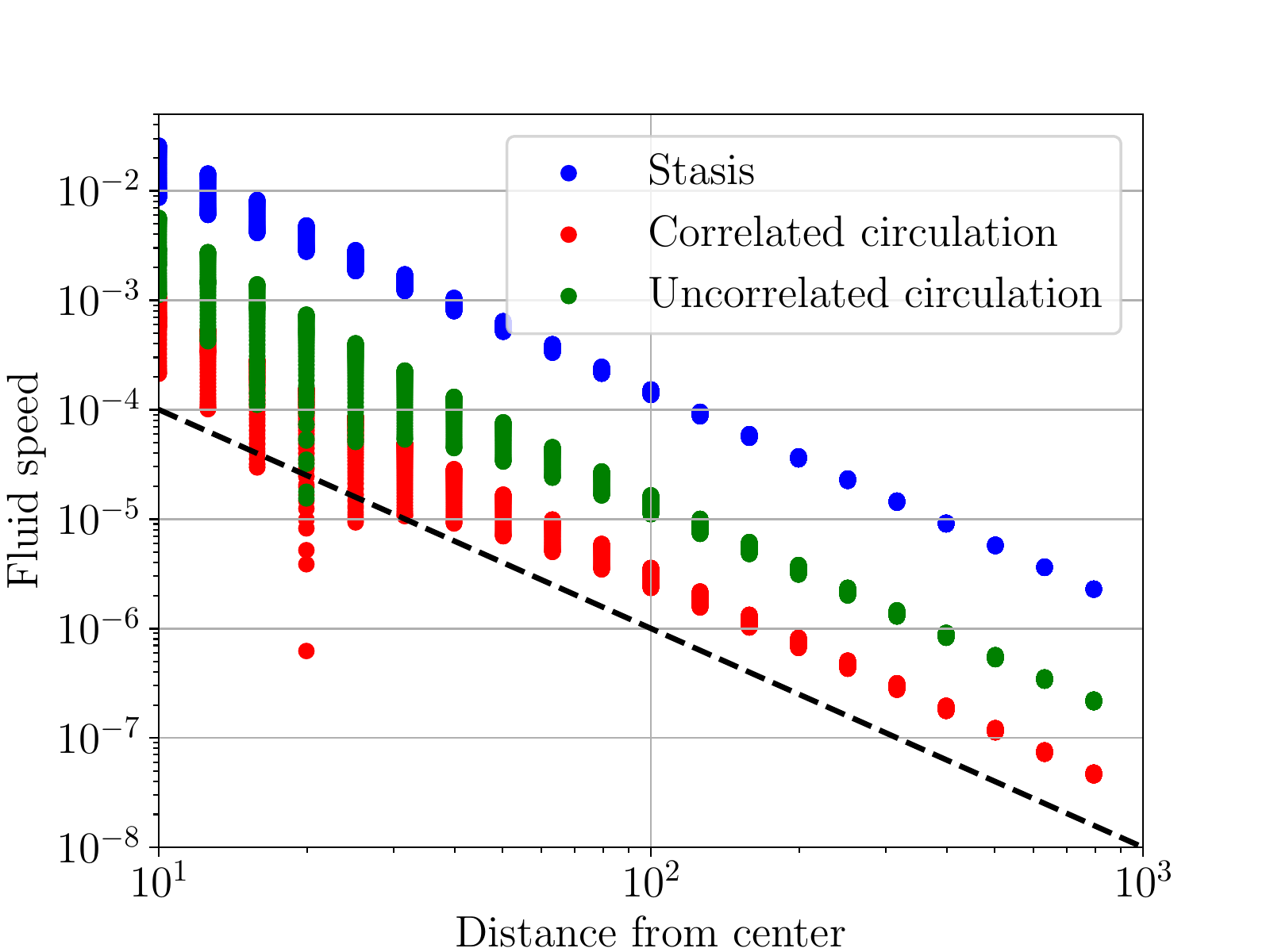}
}
\caption{Fluid speeds plotted with respect to distance to the center of the trap, with trap radius $R=5$, for the three simulations shown in Figure \ref{fig:quiver}. The black dashed line is an example of $v\propto 1/r^2$ as a guide to the eye.\label{fig:farfield}}
\end{figure}

Fig. \ref{fig:quiver} shows the direction of the velocity field for the three types of motion discussed in the previous section. We immediately notice that the velocity field for circulation behaviors is irregular, while that of stasis resembles the flow field for a stokeslet. This suggests that the far-field behavior for polymers in stasis will be stronger than those in circulation. Fig. \ref{fig:farfield} shows the average fluid speed far from the trap for the same simulations as Fig. \ref{fig:quiver}. Indeed, this is what we see: in the far field, fluid speeds due to polymers in stasis are an order of magnitude higher than polymers in uncorrelated circulation, and nearly two orders of magnitude higher than polymers in correlated circulation. Furthermore, we verify that fluid speeds for all cases drop off as $1/r^2$. This is not entirely unexpected, as this is the far-field behavior of the Liron-Mochon interaction tensor\cite{Liron1976,martin2018emergence}.

\subsection{Angular momentum power spectra}

\begin{figure*}
\centering
\subfloat[][]{
\includegraphics[width = 0.28\textwidth]{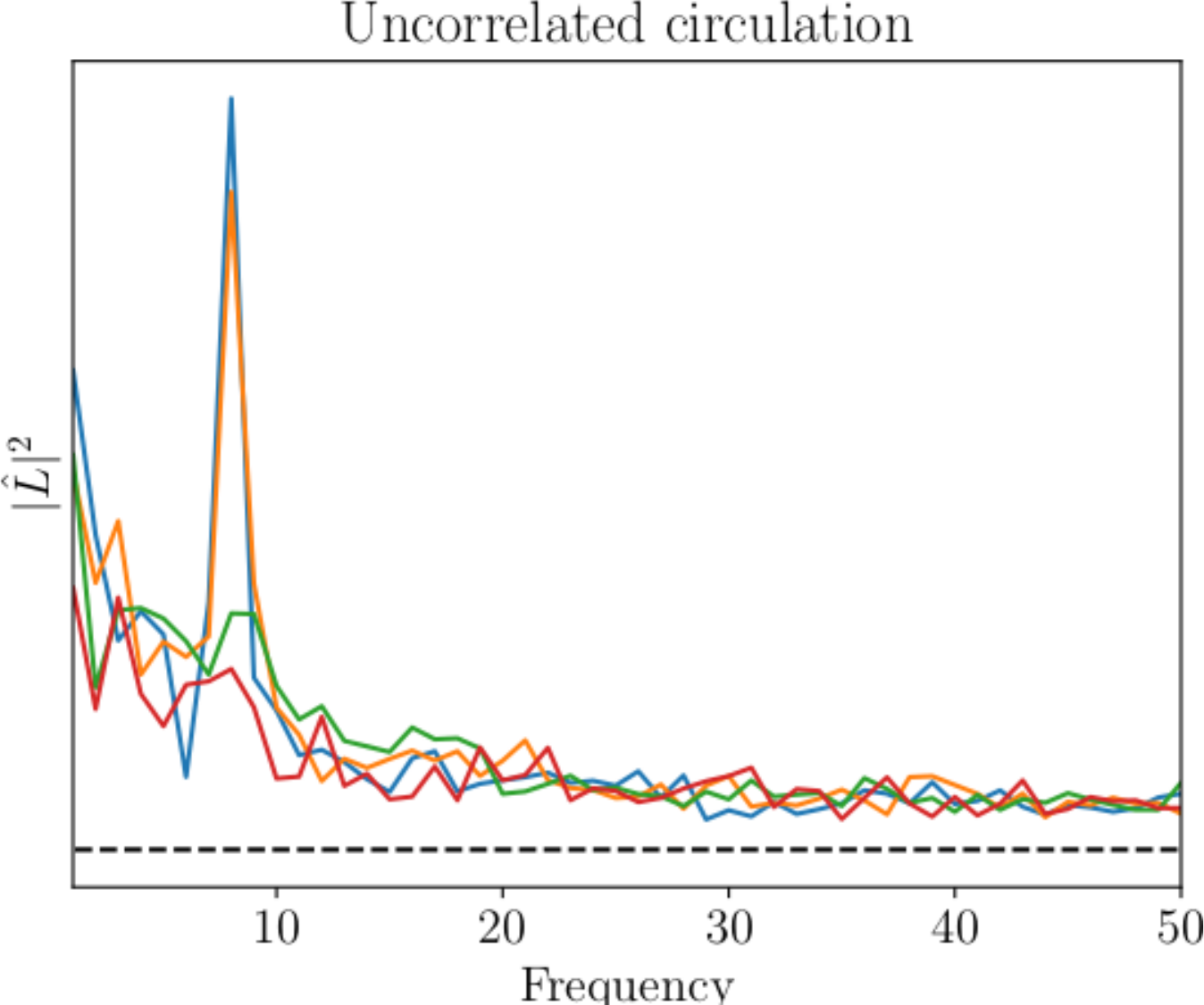}
}
\subfloat[][]{
\includegraphics[width = 0.28\textwidth]{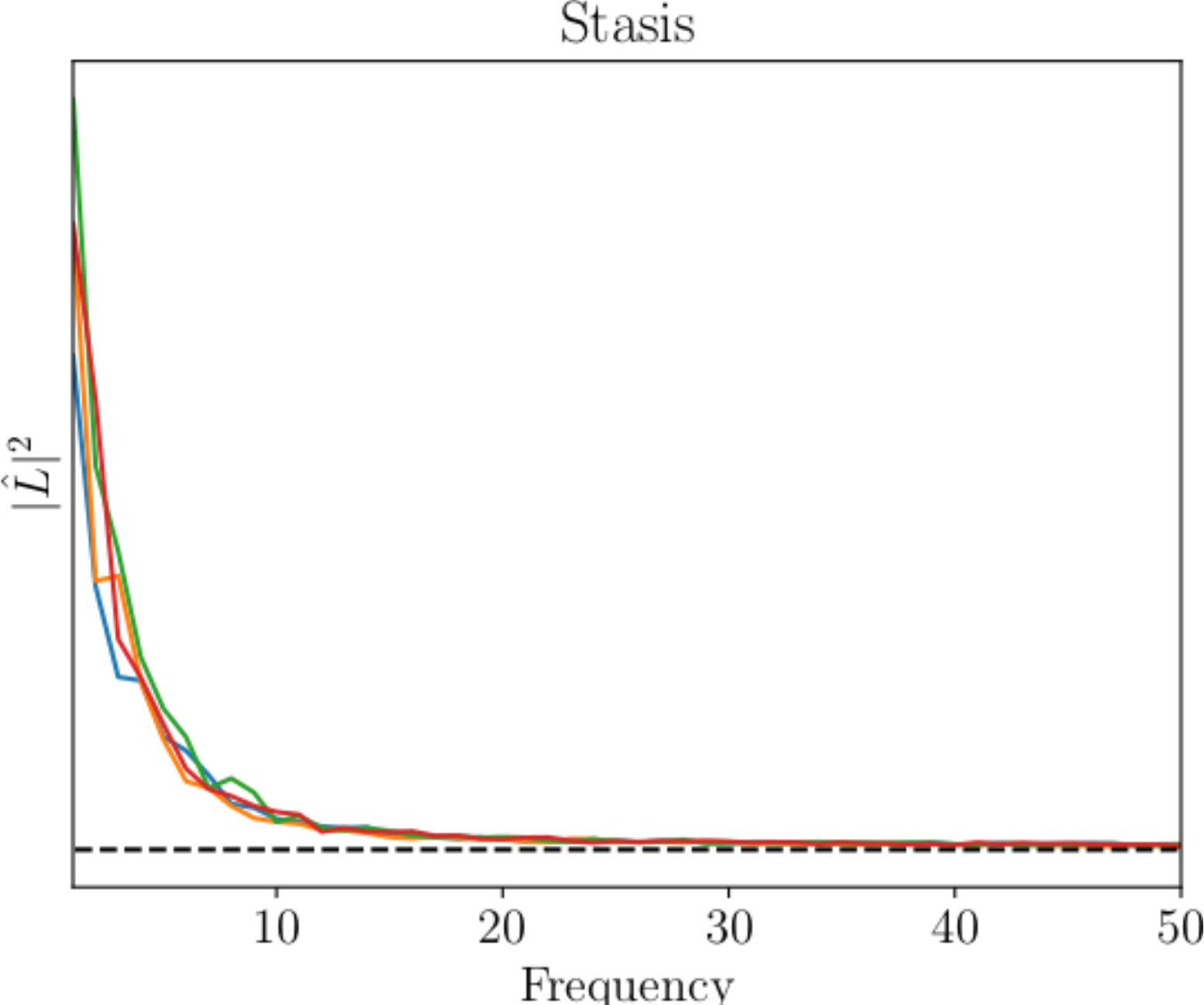}
}
\subfloat[][]{
\includegraphics[width = 0.32\textwidth]{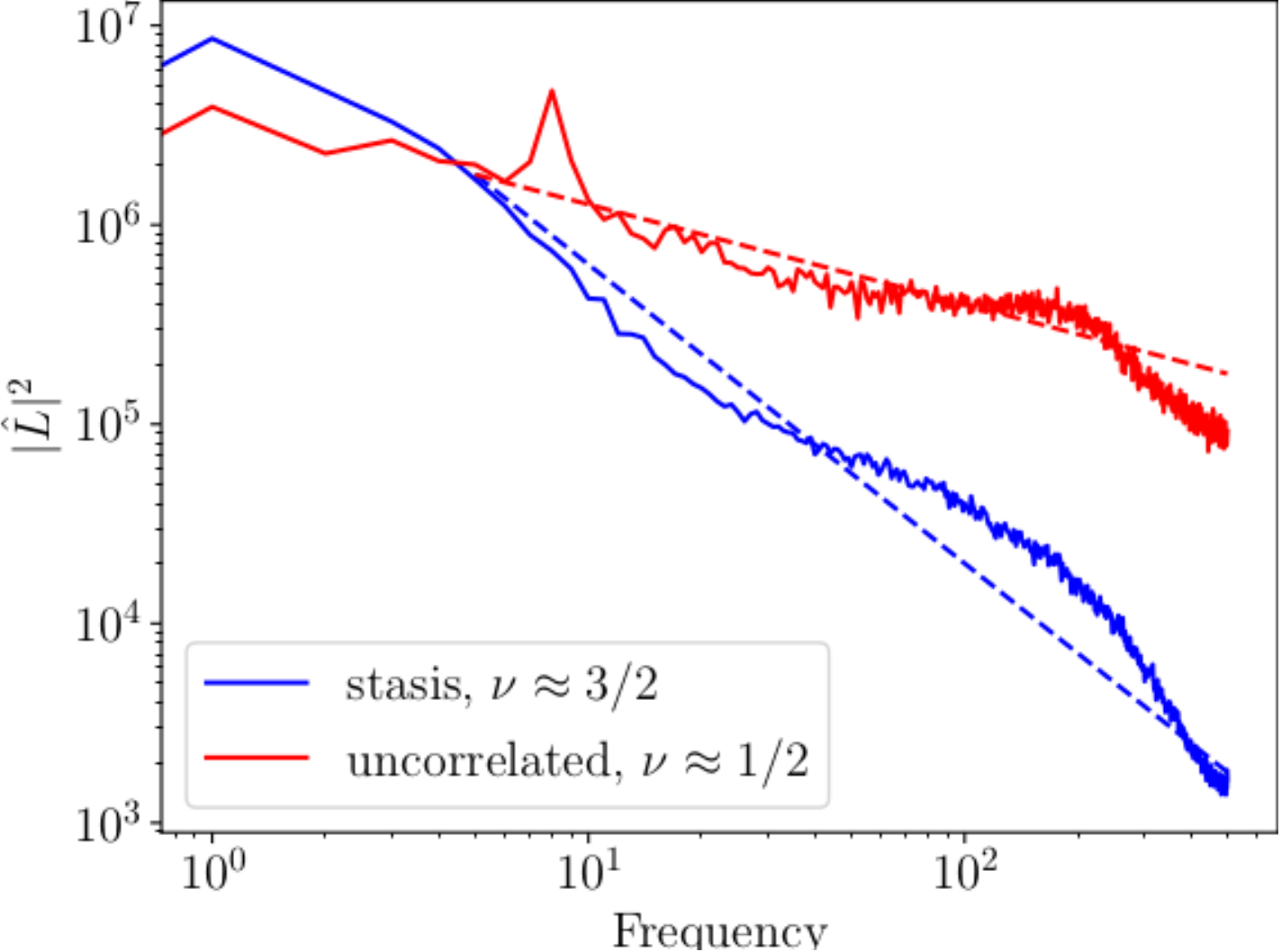}
}
\caption{Plots showing the power spectrum of (a) uncorrelated circulation and (b) stasis, with identical system parameters as shown in Fig. \ref{fig:quiver}(a) and (c), respectively. Each plot shows the power spectrum for four system initializations. Units on both axes are arbitrary. The power spectrum for correlated circulation is not shown, as it is so heavily peaked at zero frequency. Plot (c) averages the initializations from plot the other plot on a log-log scale. The dashed lines are a guide to the eye for $|\hat{L}|^2\sim \omega^{-3/2}$ (blue) and $|\hat{L}|^2\sim \omega^{-1/2}$ (red) \label{fig:powerspectrum}}
\end{figure*}

One method to quantify the differences between these types of behavior is to consider the power spectrum of the angular momentum. We first calculate the angular momentum of the $j$th monomer of the $i$th polymer,
\begin{equation}
\label{eq:L_mon}
\ell_{ij}(t) = \mathbf{r}_{ij}(t)\times \mathbf{v}_{ij}(t)
\end{equation}
Note that, due to our 2D geometry, the cross product is always in the $z$-direction and can as such be treated as a scalar. We then sum these to find the angular momentum $L_{i}$ of the $i$th polymer,
\begin{equation}
\label{eq:L_poly}
L_i(t) = \sum_{j=1}^N \ell_{ij}(t)
\end{equation}
We then find the power spectrum for each polymer over time and sum these, i.e.
\begin{equation}
\label{eq:L_fin}
|\hat{L}|^2 = \sum_{i=0}^M|\hat{L}_i|^2
\end{equation}
where $\hat{L}_i(\omega) = \mathcal{F}[L_i(t)]$ is the Fourier transform of equation \ref{eq:L_poly}. Fig. \ref{fig:powerspectrum}(a)-(b) shows this power spectrum for the uncorrelated circulation and stasis phases. The correlated circulation phase is not shown as the power spectrum is very strongly peaked at frequency $\omega=0$: that is, all polymers are more or less locked into stable circulation, and there is very little change in the angular momentum. 

We immediately notice that the uncorrelated circulation state tends to have a single peak for $\omega>0$. This makes sense, as any two given polymers circulating in opposite directions tend to collide twice per rotation, i.e. $\omega \approx \frac{2}{T}$, where $T$ is the period of circulation. The power spectra in Fig. \ref{fig:powerspectrum} were over a span of 1000 time steps (giving the $x$-axis units of $\Delta\omega = \frac{1}{1000}$), and the period of circulation for a single polymer was observed to be $T\approx200$ time steps. Indeed, we see that $\omega_{peak}\approx \frac{2}{T} = \frac{1}{100} = 10\Delta\omega$.

Fig. \ref{fig:powerspectrum}(c) shows these same spectra on a log-log scale, but each averaged over its four initializations. The intent of this plot is to show that the power spectrum for the uncorrelated circulation phase has a much larger high-frequency tail compared to that of the stasis phase. Indeed, if we assume the power spectrum has the approximate form
\begin{equation}
|\hat{L}|^2 \sim \omega^{-\nu}
\end{equation} 
then this figure shows that $\nu\approx 3/2$ and $1/2$ for the stasis phase and the uncorrelated circulation phase, respectively.

In summary, this power spectrum is a very useful tool for determining type of motion:
\begin{itemize}
\item For the correlated circulation phase, $|\hat{L}|^2$ is essentially a delta function at $\omega=0$.
\item For the uncorrelated circulation phase, $|\hat{L}|^2$ has a distinctive peak at $\omega \approx \frac2T$, where $T$ is the period of circulation. Additionally, for high frequencies, $|\hat{L}|^2\sim \omega^{-1/2}$.
\item For the stasis phase, $|\hat{L}|^2\sim \omega^{-3/2}$ for high frequencies.
\end{itemize}

\section{Conclusion}

We have shown that, under the right conditions, interesting and unique motions can occur for confined microtubules. While the design of such an optical trap experiment poses some challenges (a powerful visible-frequency laser would likely be required to contain microtubules while mitigating temperature increase), these are not prohibitive. In fact, we have shown that the trap radius should have no effect on the amount of power required to confine the microtubules, giving a fair amount of leniency in experimental design.

Three distinct types of polymer motion were identified and analyzed using first-principles simulations, with insights presented as to what parameter regimes might lead to preference of one type of motion over another. We also calculated velocity fields in the vicinity of the traps and in the far-field. These calculations are important for any future applications in mixing, as it shows that fluid motion is far more localized for correlated circular motion than for other phases (although $v\propto 1/r^2$ for all types of motion).

From an experimental perspective, it may seem as though many of the parameters varied in the simulations are not tunable. For instance, one cannot substantially vary polymer stiffness when dealing with real microtubules. However, it is possible to vary other experimental parameters to achieve the same effects. For instance, the stasis phase would more likely be encouraged with: (1) a very thin system, as this both increases the strength of hydrodynamic interactions and makes the microtubules less prone to sliding over one another, when they have crossed; (2) with more kinesin (or cargo for kinesin added) to increase viscous drag; and/or (3) increased fluid viscosity. Any increase in hydrodynamic interactions would, in effect, be equivalent to lowering the microtubule stiffness. In fact, this is one of the reasons why it is believed that Drosophila oocytes transition from the slow to fast streaming phase. Experiments \cite{serbus2005dynein} show that loosening the actin network, and hence lowering the viscosity, causes premature fast streaming. The slow streaming phase shows some similarity to the stasis phase described above (characterized by slow uncorrelated motions where the microtubules appear disordered), and fast streaming resembles correlated circulation.

Although the hydrodynamics of the boundary would be different, a related system would be to place microtubules inside a hard-walled cylinder instead of an optical trap. Because of the strong hydrodynamic screening induced by the plates, the types of behavior seen in a hard-walled experiment may be similar to that of an optical trap if the height of the cylinder is much less than its radius. It might therefore be of interest to try to confine microtubules this way as well, perhaps by forming them in situ inside of a thin circular boundary (with radius on the order of the microtubule length) that is then confined between two plates. In addition to having fewer design obstacles, such an experiment would also perhaps be more representative of intracellular systems.

The experimental confirmation of this effect would have important implications. For instance, the core forces and geometry of this work are very similar to those present in active nematics\cite{sanchez2012spontaneous,decamp2015orientational}, so understanding this behavior would be highly relevant to such systems. Such experiments could also provide a logical next step to applications in localized mixing in microfluidics.

\section{Acknowledgements}

S.E.M. was partially supported by the ARCS Foundation. 
This work was also supported by the Foundational Questions Institute \url{<http://fqxi.org>}.

\appendix

\begin{figure*}
\centering
\subfloat[][]{
\includegraphics[width=0.36\textwidth]{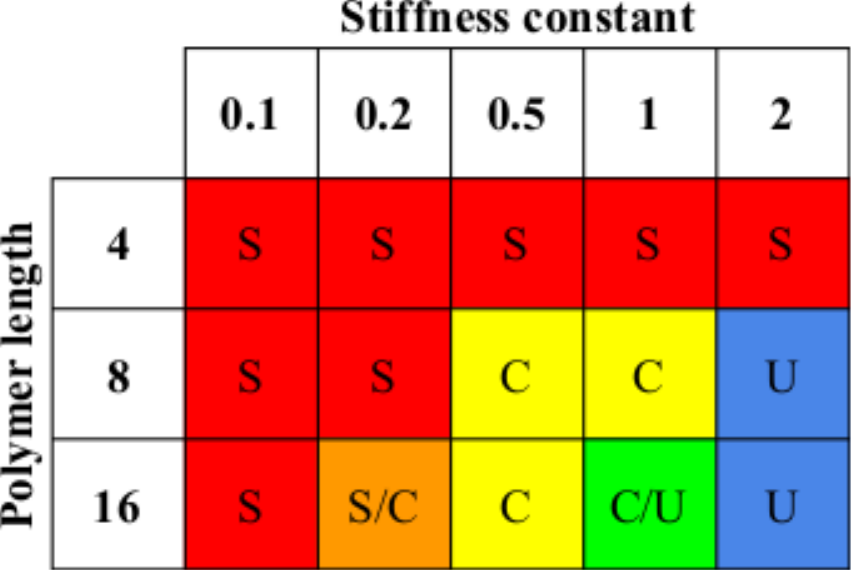}
}
\qquad\qquad
\subfloat[][]{
\includegraphics[width=0.3\textwidth]{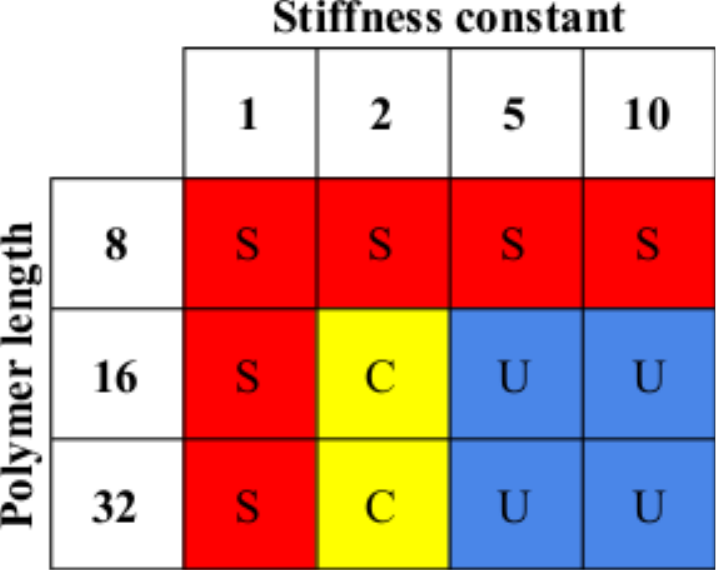}
}
\ \\
\ \\
\subfloat[][]{
\includegraphics[width=0.3\textwidth]{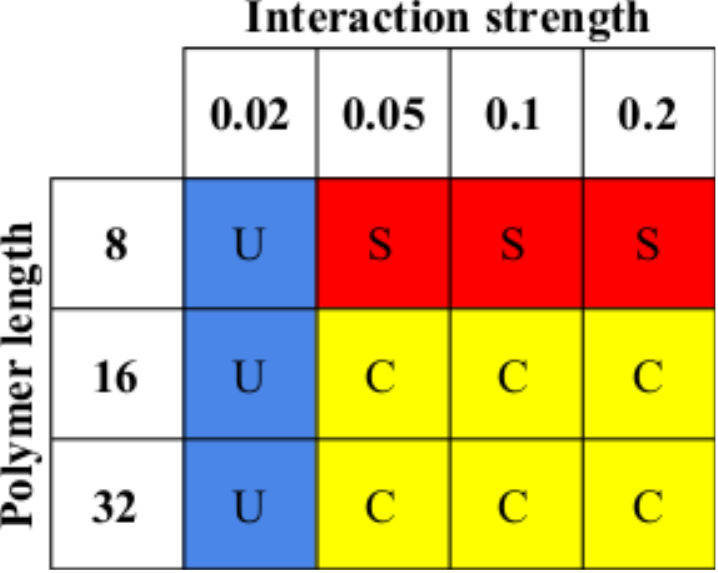}
}
\qquad\qquad\qquad
\subfloat[][]{
\includegraphics[width=0.24\textwidth]{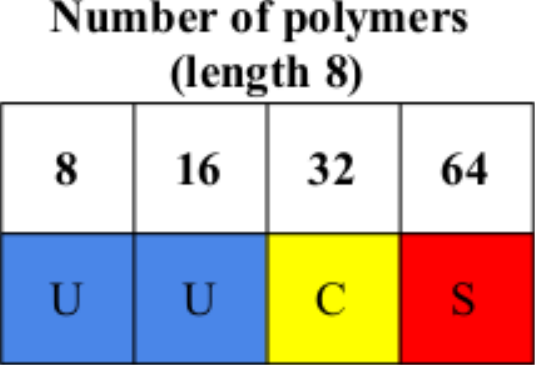}
}
\caption{Tables showing the observed behavior (S = stasis, C = correlated circulation, U = uncorrelated circulation) as a function of input parameters. In all tables except (d), monomer density remains constant -- e.g. there are double the number of polymers of length 4 as there were of length 8 in a similar run. (a) Trap radius = 5.0, 256 monomers, $k_{Oseen} = 0.1$. (b) Trap radius = 10.0, 1024 monomers, $k_{Oseen} = 0.1$. (c) Trap radius = 5.0, 256 monomers, $k_{stiff} = 0.5$. (d) Trap radius = 5.0, polymer length = 8, $k_{Oseen}=0.1$, $k_{stiff} = 0.5$. \label{fig:categories}}
\end{figure*}

\section{Videos of simulated behavior}
\label{app:vid_links}

Selected videos may be found at \url{https://sites.google.com/ucsc.edu/joshdeutsch/optical-trap-videos?authuser=0}

\section{System behavior and simulation parameters}
\label{app:categories}

Several explorations into the effects of parameter tuning are summarized in Fig. \ref{fig:categories}. The key insights of these tables are that the correlated circulation motion is sensitive to polymer density and polymer length (relative to trap radius). It is also important to take scaling concerns into account: for example, if the trap radius is doubled, the stiffness must be quadrupled in order to see analogous behavior. Simulations were also completed for a trap radius of 2.5, but correlated circulation was never seen for this radius (likely because the expected polymer length for circulation is now only 4, which does a poor job approximating an elastic rod).

\clearpage

\bibliography{Opt_Trap}

\end{document}